\documentclass{aa}
\usepackage{graphicx}
\usepackage{natbib}

\begin{document}

\title{Expanding shells of shocked neutral hydrogen around compact \ion{H}{ii} regions}

\author{Roland Kothes \inst{1,2} \and Charles R. Kerton \inst{1}}

\authorrunning{Kothes \& Kerton}

\titlerunning{Shocked \ion{H}{i} shells around compact \ion{H}{ii} regions}

\offprints{R. Kothes}

   \institute{National Research Council of Canada,
              Herzberg Institute of Astrophysics,
              Dominion Radio Astrophysical Observatory,
              P.O. Box 248, Penticton, British Columbia, V2A 6K3, Canada\\
         \and
             Department of Physics and Astronomy,
             The University of Calgary, 2500 University Dr. NW,
             Calgary, AB, T2N 1N4 Canada\\
\email{roland.kothes@nrc.ca, charles.kerton@nrc.ca}}

\date{Received; accepted }

\abstract{

By comparing radial velocities of radio bright
compact \ion{H}{ii} regions with their \ion{H}{i} absorption profiles, 
we discovered expanding shells of neutral hydrogen around them. 
These shells are revealed by
absorption of the radio continuum emission from the HII regions at 
velocities indicating greater distances than the observed 
radial velocity. We believe that these shells are shock
zones at the outer edge of the expanding ionized region. Additionally 
we found evidence for a velocity inversion inside the Perseus arm
caused by a spiral shock, which results in a deep absorption
line in the spectra of compact \ion{H}{ii} regions behind it.

 -- ISM: kinematics and dynamics}

\maketitle

\section{Introduction}

Models of the photodissociation of molecular clouds surrounding newly
formed O and early B type stars \citep{hh78,rog92} show
that the resulting \ion{H}{i} will exist in a layered structure
surrounding the star.  A dissociation front (DF) will first rapidly
move through the molecular cloud forming a broad \ion{H}{i} region. As
the \ion{H}{ii} region expands a layer of shocked \ion{H}{i} will form
just outside the ionization front (IF).  Eventually the
faster moving IF will catch up with the DF and all of the \ion{H}{i}
will be found in the shocked layer. The time when the IF and DF merge
ranges from $\sim 10^5 - 10^6$ years depending upon the spectral type
of the star and the density of the surrounding molecular material 
\citep[see Figure 7 of][]{rog92}. The broad \ion{H}{i} region  has been
detected in 21-cm line emission in a number of studies 
\citep[e.g. ][]{rog82}, but observational evidence for the shocked layer 
is currently limited to a study of the \ion{H}{ii} region Orion A
by \citet{vdw89}.

In this paper we present a novel technique for the detection and study of
expanding \ion{H}{i} structures around compact \ion{H}{ii} regions that
uses 21-cm \ion{H}{i} absorption spectra. The technique allows \ion{H}{i} 
surrounding \ion{H}{ii} regions to be studied even if the \ion{H}{ii} region 
is not resolved. In a number of cases we detect absorption features that are
associated with shocked expanding layers of photodissociated gas. 
We focus on compact \ion{H}{ii} regions within the observed area of the
Canadian Galactic Plane Survey (CGPS)\footnote{CGPS data are
publicly available via the Canadian Astronomy Data Centre (CADC; http://cadcwww.
hia.nrc.ca)} ($74.2\degr \le l \le 147.3\degr$; $-3.6\degr \le b \le 5.6\degr$).
The CGPS is part of the current international effort to
map the Galactic plane in the 21-cm \ion{H}{i} line at $\sim 1\arcmin$ 
resolution \citep{tay99, mcc01}.  At this resolution many Galactic
\ion{H}{ii} regions will remain unresolved and the surrounding neutral
material must be studied via absorption. The technique presented in
this paper, applied to such regions, will greatly increase the number of 
\ion{H}{ii} regions studied using 21-cm absorption spectra.

\ion{H}{i} absorption spectra towards radio bright Galactic \ion{H}{ii}
regions have been used primarily to probe the properties of the
neutral ISM along the line of sight to the \ion{H}{ii} region \citep{wen96, 
nor99}. In these studies absorption features at velocities associated
with the \ion{H}{ii} region itself are usually ignored. Another
application has been in  Galactic structure studies; \citet{kb94} and 
\citet{kb90} have shown how 21-cm absorption spectra can resolve the
distance ambiguity towards inner galaxy \ion{H}{ii} regions.

The \ion{H}{i} absorption spectrum towards a galactic \ion{H}{ii} region can
also be used to study the kinematics and structure of neutral material
associated with the \ion{H}{ii} region.  Such absorption studies
have tended to be limited to a few very bright \ion{H}{ii} regions such as DR
21 \citep{rob97}, Orion \citep{vdw89} and W3 \citep{vdw90}. More
recently \citet{le01} used \ion{H}{i} absorption as part of a
multiwavelength study of the blister \ion{H}{ii} region G111.61+0.37.

In \S 2 we describe the data used and outline the data processing techniques.
Particularly we focus on a new spatial structure filtering technique. In
\S 3 we show the absorption spectra of all our
\ion{H}{ii} regions.  A general model for the structure of an absorption
spectrum is presented and applied to our observations in \S 4 followed
by a general discussion in \S 5.
Finally, a summary is presented in \S 6.

\section{Observations and Data Analysis}

\subsection{\ion{H}{i} Line Data from the CGPS}

\ion{H}{i} 21-cm line data were obtained using the DRAO synthesis telescope
\citep{lan00} as part of the Canadian Galactic Plane Survey (CGPS)
\citep{tay99}. The \ion{H}{i} data reduction techniques used in the CGPS are
described in detail in \citet{hig00}. The CGPS consists of data cubes with
a size of $5.12\degr \times 5.12\degr$. For further
analysis we extracted $1.4\degr \times 1.4\degr$ fields for each
\ion{H}{ii} region studied. 

\subsection{Data Analysis}

To determine \ion{H}{i} emission and absorption profiles, we have to separate
the Galactic background emission from the small scale structures
around and on the compact \ion{H}{ii} regions we plan to study. 
For that purpose we used the ``background filtering technique'' invented 
by \cite{sofue}. This method was developed for separating
smooth Galactic background emission from small scale emission
structures. We have modified the technique
to be useful for negative features as well as for positive
ones. 

For each velocity channel we used the following procedure:
\begin{itemize}

\item $T_B \rightarrow T_B^0$ convolution to a beam $\Theta$

\item $T_B - T_B^0 \rightarrow \delta T_B$

\item $T_B \times \delta T_B \rightarrow \Pi T_B$

\item $T_B^1 = \left\{
               \begin{array}{lcl}
               T_B & : & \Pi T_B < 0 \\
               \delta T_B & : & \Pi T_B \ge 0 \\
               \end{array}\right\}$
               
\item $T_B^1 \rightarrow T_{large}$ convolution to $\Theta$

\item $T_B - T_{large} \rightarrow T_{small}$

\end{itemize}

Here $T_B$ is the original brightness temperature distribution with
subtracted continuum emission, $T_{large}$ 
represents the large scale background emission, and
$T_{small}$ the small scale structures including the
absorption of the compact \ion{H}{ii} regions.

For our \ion{H}{i} data
this technique creates one \ion{H}{i} data cube representing large scale
structures which provides information about the emission of the
smooth background component for each source. We also get one \ion{H}{i} data
cube containing small scale structures. Absorption profiles towards the 
compact \ion{H}{ii} regions and emission profiles of possible \ion{H}{i}
structures around them are derived from this cube.


\section{Absorption profiles of compact \ion{H}{ii} regions}

\subsection{The source sample}

\begin{table}
{
\begin{tabular}{c|r@{$\pm$}lrc|l} \hline
   Name & \multicolumn{2}{c}{$T_c^e$} & v$_{LSR}$ &  Reference &  other names\\
    & \multicolumn{2}{c}{[K]} & [km/s] &  & \\ \hline
    G75.77+0.35 & 450 & 23 & -8.5 & (1)  & \\
    G75.83+0.40 & 750 & 38 & -4.1 & (1)   & Sh 105 \\
    G76.15-0.29 & 100 & 8 & -28.2 & (1)   & \\
    G76.19+0.10 & 55 & 6 & -2.1 & (2)   & IRAS 20220+3728 \\
    G80.35+0.73 & 100 & 11 & -62.4 & (1)   & \\
    G80.94-0.13 & 374 & 21 & -2.1 & (1)  & IRAS 20375+4109 \\
    G83.94+0.77 & 143 & 9 & -8.5 & (1)   & \\
    G85.25+0.02 & 69 & 7 & -31.2 & (1)   & \\
    G85.69+2.03 & 41 & 5 & -75.0 & (4)   & IRAS 20446+4613 \\
    G90.24+1.72 & 32 & 5 & -60.9 & (3)   & Sh 121 \\
    G92.69+3.08 & 48 & 6 & -6.4 & (2)  & IRAS 21078+5211 \\
    G93.86+1.56 & 44 & 5 & -59.5 & (2)   & IRAS 21202+5157 \\
    G96.29+2.60 & 52 & 5 & -94.7 & (3)   & Sh 127 \\
    G97.51+3.17 & 72 & 5 & -71.1 & (2)  & Sh 128 \\
    G105.62+0.34 & 71 & 5 & -52.1 & (2)  & Sh 138 \\
    G108.20+0.58 & 259 & 14 & -55.9 & (1)   & Sh 146 \\
    G108.37-1.06 & 58 & 5 & -53.1 & (3)  & Sh 149 \\
    G108.76-0.95 & 150 & 9 & -49.1 & (1)  & Sh 152 \\
    G110.11+0.05 & 225 & 12 & -51.7 & (1)   & Sh 156 \\
    G111.62+0.37 & 87 & 6 & -63.4 & (1)   & Sh 159 \\
    G112.23+0.24 & 82 & 7 & -41.8 & (1)  & Sh 162 \\
    G114.02-1.45 & 63 & 5 & -49.0 & (5)   & IRAS 23365+5953 \\
    G115.79-1.58 & 33 & 4 & -40.8 & (1)  & Sh 168 \\
    G132.16-0.73 & 79 & 5 & -56.6 & (1)   & IRAS 02044+6031 \\
    G137.75+3.80 & 77 & 5 & -47.0 & (5)   & IRAS 03030+6229 \\
    G138.49+1.64 & 61 & 5 & -34.6 & (1)   & Sh 201 \\ \hline
    \end{tabular}}
\caption{Characteristics of the compact \ion{H}{ii} regions in our
sample. $T_c^e$ represents the peak continuum brightness temperature
in the CGPS 1420~MHz data and v$_{LSR}$ the radial velocity. References
for the radial velocities:
(1) \citet{lockman}, (2) \citet{bronfman}, (3) \citet{brand}, 
(4) \citet{wouterloot}, (5) \citet{brunt}. Ref. (1) contains
recombination line velocities and all other references 
radial velocities of molecular lines.}
\label{hiidata}
\end{table}

We obtained absorption profiles towards all \ion{H}{ii} regions
in the CGPS survey area which are unresolved in the 1420~MHz continuum
data and have a flux density greater than 200~mJy (T$_b \sim 30$~K) therein. 
We obtained radial velocities for these \ion{H}{ii}
regions from various sources as listed in Table \ref{hiidata}. 
Sources with positive radial velocities were removed from the sample 
because of the 
distance ambiguity in the inner Galaxy. All sources in our sample
are bright enough in the CGPS to show significant absorption by foreground
neutral hydrogen. 

\subsection{Obtained absorption profiles}

We used the background filtering technique described in Sect. 2.2 to
separate the broad structures from the compact absorption features.
We used a beam $\Theta$ of $5\arcmin$ for all sources and checked all final
data cubes carefully by eye. For three stronger sources,G75.77, G75.83, 
and G80.94, there were still some
faint features from the absorption visible in the background component.
For these we used a beam of $10\arcmin$ which then successfully separated  
small scale and large scale structures. We created the absorption spectra in 
Figs. \ref{firstspec} to \ref{lastspec} by simply plotting the value at 
the peak position of the radio continuum source in each \ion{H}{i} channel of 
the \ion{H}{i} cubes as a function of the radial velocity. 
One naively expects to see 
absorption of the \ion{H}{ii} region's continuum emission by the neutral 
hydrogen in the foreground up to its radial velocity maybe with a Gaussian 
slope with a width of a few km/s towards higher negative velocities,
assuming a flat rotation model. In contrast
we see at least one additional absorption line well beyond the radial 
velocity in all but three of the 26 observed \ion{H}{ii} regions. 
The additional absorption lines are at higher negative velocity than the
radial velocity of the source itself indicating material which is moving 
towards us relative to the \ion{H}{ii} region. Considering the fact that we
find this feature in almost all of the observed compact \ion{H}{ii} regions, 
we can assume that it is a symmetric expanding shell of neutral hydrogen.
Since we are dealing with compact \ion{H}{ii} regions which are unresolved
in our data we do not find those expanding shells around them in emission
because the emission is buried together with the foreground
absorption within the beam. 

We should note at this point that the additional absorption line could also
be caused by a double valued rotation curve. Our sample was chosen to
avoid the velocity ambiguity in the inner Galaxy, but there is also
evidence for a velocity inversion in the Perseus arm caused by a spiral 
shock \citep{roberts}. However, since we
observe this phenomenon for almost all of the compact \ion{H}{ii} regions
independent of the location and distance we assume this to be a minor factor.
Nevertheless, we will discuss this in more detail in Sect. 5.

\section{Theory and Modelling}

\subsection{Dissociated molecular gas around massive young stars}

In this section we summarize some of the basic results of the models of
\citet{hh78} and \citet{rog92} regarding the morphology, velocity,
temperature and column density structure of the dissociated gas surrounding a
newly formed OB star.  Essentially these models demonstrate that the
dissociated gas will have a layered structure consisting of a broad
unshocked layer formed by a dissociation front (DF) and a thinner
shocked layer of \ion{H}{i} located just outside of the expanding
ionization front (IF). 

The shocked layer of \ion{H}{i} initially expands with the
\ion{H}{ii} region at velocities $\sim$ 10 km/s. For extremely young systems, 
where there has been little or no
expansion of the \ion{H}{ii} region, the shocked layer of \ion{H}{i}
will not be present \citep[e.g. ][]{dr82}. By
the time IF merges with the more slowly moving DF the expansion
velocity of the shocked layer will have dropped to $\sim 3-4$~km/s. The merging
of the two fronts occurs on a time scale of $10^5 - 10^6$ years
depending upon the density of the molecular material and the spectral
type of the exciting star. The unshocked \ion{H}{i} layer is not
expected to have any systematic velocity shift relative to the molecular 
material if it is
fully confined by the molecular material. Observations that do show 
expansion of the unshocked dissociated \ion{H}{i} are all associated
with blister \ion{H}{ii} regions \citep{rog82,vdw89}. The
observed expansion velocities of 2-6 km/s in these cases are most
likely due to the expansion of the heated dissociated gas into the
surrounding low density medium.

Gas temperatures are expected to be $ \leq 100$~K within the
unshocked material; newly shocked material may be hotter but is
expected to cool rapidly to T$ = \mathrm{few} \times 100$~K  
because of efficient cooling in the shocked layer \citep{rog92}. 
Observational support for these values come from {\it ISO} observations 
of H$_2$ lines towards photodissociation regions around S 140 \citep{tim96} 
and S 106 \citep{vda00} which indicate T$\sim 500$~K, and from 
\ion{H}{i} 21-cm line emission observations of a compact 
\ion{H}{ii} region in the nebular complex GGD 12-15 which indicate 
T$\sim 300$~K \citep{gom98}.

The column density of shocked \ion{H}{i} is expected to grow as the system
evolves, and has typical values  of $10^{20} - 10^{21}$ over the
lifetime of the system \citep{rog92}. The unshocked
layer is expected to reach a peak column density of $\sim 1-3\times
10^{21}$ with little dependence on the spectral type of the exciting
star and the density of the molecular material \citep{rog92}. 
  
\subsection{A simple model for expanding \ion{H}{i} shells}
 
To calculate observational properties of the expanding \ion{H}{i} shells
around the compact \ion{H}{ii} regions we have
developed a simple model based upon \citet{hh78} and
\citet{rog92}. An early type star, located in the centre of
a molecular cloud, creates an expanding
\ion{H}{ii} region inside an area of dissociated molecular material
(see Fig.~\ref{model}). At the interface we have a narrow zone of shocked
neutral material expanding together with the ionization front.
For simplicity we assume the whole structure to be symmetric. 
The object is located at a distance $d$ from
the Sun which translates to a radial velocity $v_d$ assuming a flat
rotation curve with $v_\odot$ = 220~km/s and $R_\odot$ = 8.5~kpc. 
We further assume that the object is located
in the outer Galaxy so that the velocity $v_d$ is negative and a higher
negative velocity implies a greater distance from the Sun. 
The part of the shocked \ion{H}{i} shell moving towards us would be 
shifted in its radial velocity to $v_d - v_{exp}$, with $v_{exp}$ 
representing the expansion velocity of the \ion{H}{i} shell, while the 
velocity of the receding shell would be shifted to $v_d + v_{exp}$ 
(see Fig. \ref{model}). In this case the receding \ion{H}{i}, even though 
it is located behind the \ion{H}{ii} region, would be observed among 
Galactic \ion{H}{i} located between the object and the Sun. The 
approaching part of the shell, even though located
in front of the \ion{H}{ii} region, would appear among the material which is
located behind the object in our \ion{H}{i} data (see Fig. \ref{model}).

Assuming a constant spin temperature $T_S$ within the shocked neutral
material the observed brightness temperature $T_B$ at the position of the 
\ion{H}{ii} region would be:
\begin{equation}
   T_B(v) = T_S^e (1-e^{-\tau(v)})
\end{equation}
for the receding part and:
\begin{equation}
   T_B(v) = T_S^e (1-e^{-\tau(v)}) + T_C^e(e^{-\tau(v)} -1)
   \label{tb}
\end{equation}
for the approaching part, where the second term represents the 
absorption of the \ion{H}{ii} region's continuum emission by the 
approaching \ion{H}{i} shell. Here $T_C^e$ is the \ion{H}{ii} region's 
effective continuum brightness temperature in our 1420~MHz data and $\tau$ 
is the optical depth within the \ion{H}{i} shell. $T_S^e$ is the
effective spin temperature. Since these compact \ion{H}{ii} regions are
unresolved in our data the peak brightness temperature is not the real
brightness temperature of the source. 
$T_C^e$ is $T_C$ multiplied by the ratio of the actual source area
to the observed source area, that is:
\begin{displaymath}
   T_C^e = T_C\frac{\Theta_S^2}{\Theta_B^2 + \Theta_S^2},
\end{displaymath}
where $\Theta_B$ is the HPBW of the observation and
$\Theta_S$ the angular diameter of the source. For a source with a diameter of
$10\arcsec$ and a brightness temperature of 1000~K for example we 
would measure a peak
of only 30~K. The same effect applies to the spin temperature.

If we now create \ion{H}{i} absorption line spectra as described in
Sect. 3.2 we will find continuum absorption of the \ion{H}{ii} regions
radio emission by the foreground material up to its radial velocity. Since we
find the receding part of the expanding \ion{H}{i} shell at a radial
velocity of $v_d + v_{exp}$ it will appear as an emission line superimposed
on the absorption structure of the foreground material making it
hard to detect. The approaching part, however,  has a radial velocity of
$v_d - v_{exp}$. We expect to find its related line
separated from the foreground absorption at a higher negative velocity.

Since our sample consists of sources which are significantly smaller
than the beam of the observations, we can only speculate about the 
structure of neutral material around these compact \ion{H}{ii} regions.
We can resolve different layers with different expansion velocities
but we do not get any information about the spatial distribution. Our
method favours the part of the expanding shells which sees most of
the \ion{H}{ii} region's radio continuum emission. In this case small
high density knots are smoothed out and thus not detectable.

\subsection{Application to the data}

We can now calculate characteristics of the expanding \ion{H}{i} shells
by fitting Gaussians to the individual absorption spectra. Four examples
are shown in Fig. \ref{gauss}. A comparison
of the \ion{H}{ii} region's radial velocity with the position of
the additional absorption line gives the expansion velocity of
the \ion{H}{i} shell. By transforming eqn. \ref{tb} we obtain
an expression for the opacity $\tau$ within the expanding shell:
\begin{equation}
    \tau = - ln \left ( \frac{T_B}{T_C^e - T_S^e} + 1 \right )
    \label{tau}
\end{equation}

\begin{table}
{
\begin{tabular}{c|r@{$\pm$}lccc} \hline
   Name &  \multicolumn{2}{c} {v$_{exp}$} & $\tau$ & $\tau$          & HPBW \\
        &  \multicolumn{2}{c}  {[km/s]}   &  & $T_s/T_c = 0.2$ & [km/s] \\ \hline
    G75.77+0.35   & 19.2 & 0.6 & $0.17\pm 0.01$ & 0.21 & 1.7  \\
    G75.83+0.40   & 25.6 & 0.7 & $0.05\pm 0.01$ & 0.07 & 1.7  \\
    G76.15-0.29   & 7.7 & 3.2 & $0.45^{+0.06}_{-0.05}$ & 0.61 & 4.9  \\
    G76.19+0.10   & 5.1 & 0.8 & $0.79^{+0.17}_{-0.13}$ & 1.15 & 2.8   \\
    G80.35+0.73   & 5.4 & 2.9 & $0.64^{+0.13}_{-0.10}$ & 0.89 & 5.1  \\
    G80.94-0.13  & \multicolumn{2}{c} {2.7?} & - & - & -   \\
    G83.94+0.77  & 11.1 & 2.2 & $0.28^{+0.03}_{-0.02}$ & 0.37 & 5.6  \\
    G85.25+0.02   & 10.8 & 2.4 & $0.49^{+0.08}_{-0.06}$  & 0.66 & 2.4  \\
    G85.25+0.02   & 5.9 & 2.4 & $0.23\pm 0.03$ & 0.30 & 1.2  \\
    G85.69+2.03  & 2.5 & 0.5 & $0.26^{+0.05}_{-0.04}$ & 0.34 & 3.8   \\
    G90.24+1.72   & 23.0 & 0.5 & $0.91^{+0.33}_{-0.20}$ & 1.37 & 2.4   \\
    G92.69+3.08  & 0.8 & 0.8 & $>$ 2.00 & $>$ 2.00 & 4.2   \\
    G93.86+1.56   & 11.0 & 0.8 & $0.82^{+0.19}_{-0.14}$ & 1.21 & 6.8  \\
    G93.86+1.56   & 2.2 & 0.8 & $0.46^{+0.08}_{-0.07}$ & 0.62 & 2.9   \\
    G96.29+2.60   & \multicolumn{2}{c}- & - & - & -   \\
    G97.51+3.17  & 9.4 & 0.8 & $0.70^{+0.10}_{-0.08}$ & 1.00 & 4.2  \\
    G97.51+3.17  & 2.2 & 0.8 & $1.60^{+0.41}_{-0.29}$ &  $>$2.00 & 3.5  \\
    G105.62+0.34  & 4.0 & 0.8 & $0.28\pm 0.03$ & 0.36 & 1.2   \\
    G105.62+0.34  & 1.5 & 0.8 & $0.68^{+0.09}_{-0.08}$ &  0.97 & 4.2   \\
    G108.20+0.58   & \multicolumn{2}{c}- & - & - & -   \\
    G108.37-1.06   & 0.6 & 1.3 & $>$ 2.00 & $>$ 2.00 & 3.6  \\
    G108.76-0.95   & 4.3 & 0.9 & $1.76^{+0.46}_{-0.32}$ & $>$ 2.00 & 5.8   \\
    G110.11+0.05   & 6.2 & 1.2 & $0.78^{+0.09}_{-0.08}$ & 1.13 & 3.2   \\
    G111.62+0.37   & \multicolumn{2}{c}- & - & - & -  \\
    G112.23+0.24   & 9.4 & 1.5 & $0.16\pm 0.02$ & 0.21 & 1.0   \\
    G114.02-1.45   & 5.4 & 0.5 & $0.89^{+0.15}_{-0.12}$ & 1.33 & 1.7  \\
    G115.79-1.58   & 4.0 & 1.5 & $1.46^{+0.63}_{-0.34}$ & $>$ 2.00 & 1.6   \\
    G132.16-0.73   & 2.5 & 4.8 & $1.54^{+0.35}_{-0.26}$ & $>$ 2.00 & 3.7   \\
    G137.75+3.80   & 6.1 & 0.5 & $0.51^{+0.06}_{-0.05}$ & 0.70 & 1.3   \\
    G138.49+1.64   & 4.3 & 1.0 & $>$ 2.00 & $>$ 2.00 & 3.1   \\ \hline
    \end{tabular}}
\caption{Derived parameters for the compact \ion{H}{ii} regions in our
sample. Sources with two additional absorption lines are noted twice.}
\label{hiidata2}
\end{table}

We do not have any information about the spin temperature within
the expanding \ion{H}{i} shell since its emission is superimposed
on the absorption profile. Assuming that the spin temperature is 
significantly lower than the continuum brightness temperature of the
\ion{H}{ii} region we can neglect it in eqn. \ref{tau} and calculate a value
for $\tau$. We should note at this point that the effect of $T_S$ on the 
resulting value of $\tau$ increases with high $T_B$ to $T_C^e$ ratios. As an
example we also calculated the resulting $\tau$ when the spin temperature is
20~\% of the continuum temperature. The results for the individual
\ion{H}{ii} regions are listed in Table \ref{hiidata2}. For $\tau>2.0$
we have indicated a lower limit since the quantity 
$\frac{T_B}{T_C^e - T_S^e}$ is smaller than our uncertainties in this case.
Three of the sources do not have additional absorption lines at higher
negative velocities and we could not obtain a satisfactory multiple
Gaussian fit to the absorption profile of G80.94 due to its complex
structure.

\section{Discussion}
 
\subsection{The nature of the additional absorption line(s)}

The discussion about the evolution of the layer of shocked neutral
hydrogen expanding with the ionizing shock front around young stellar objects 
in Sect. 4.1 gives us constraints
on the characteristics of these features. We find a maximum expansion
velocity of about 10~km/s. An upper limit of about 1.0 for the opacity
$\tau$ can be deduced from the maximum \ion{H}{i} column density, 
$N_H=10^{21}$~cm$^{-2}$, and the typical spin temperature of a few 
100~K by using the relation:
\begin{equation}
   N_{\ion{H}{i}} = 1.823\cdot 10^{18}\cdot T_S\cdot \int \tau(v) dv
\end{equation}
with
\begin{displaymath}
   \int \tau(v) dv = 1.06\cdot \tau \cdot \delta v
\end{displaymath}
for a Gaussian distribution.
Here $\delta v$ is the HPBW of the absorption line, which usually lies
around a few km/s and $\tau$ is the opacity
at the peak of the line as calculated in Table~\ref{hiidata2}.
For an examination of those constraints within our data we plotted
the opacity $\tau$ as a function of the expansion velocity for all
available sources in Fig.~\ref{tauvexp}. For those sources where we found
double structures we plotted both results to allow for multiple lines
of different origin. We found three well separated groups
of $\tau-v_{exp}$ pairs in the diagram.
Most sources are located in the lower left corner. Their $\tau$ and
$v_{exp}$ values agree very well with the constraints we laid out for 
the expanding layers of shocked \ion{H}{i}. There seems to be a sharp 
upper limit
for the expansion velocity at about 11~km/s and an upper limit for $\tau$
of about 1.0. 

There is another group of sources with low $v_{exp}$ and very high
$\tau$. We should note at this point again that
the $\tau$ values in this diagram were calculated by assuming that
$T_S << T_C$. The effect of $T_S$ on the final $\tau$ is bigger
for $\tau$ values which are already high by assuming $T_S << T_C$
(see Table\ref{hiidata2}). Hence the sources located in the lower
left corner are not that much affected by higher $T_S$ values than
those which are already high above this area. This means that we cannot explain
the sources located high above $\tau=1$ with expanding layers of shocked
\ion{H}{i}. 

We believe that the absorption features we are detecting in
the second group indicate a possible
distance ambiguity in the Perseus spiral arm. According to \citet{roberts}
the spiral shock causes the radial velocity to drop by 20 - 30 km/s at 
the beginning of the Perseus arm and slowly rejoins the flat rotation
curve after about 1~kpc. This creates a velocity inversion inside
the Perseus arm. Thus inside the arm the distance is increasing with
increasing radial velocity as opposed to the flat rotation curve. According
to \citet{roberts} the density in the closer part of the Perseus arm
is significantly enhanced by the spiral shock which would explain the
high $\tau$ values we obtained. For this interpretation we would
expect a strong
dependence of the radial velocity of this absorption line on 
Galactic longitude. Because of the changing viewing angle to
the Perseus arm the radial velocity should increase with increasing
Galactic longitude. To examine this dependence we plotted the $v_{LSR}$
at the center of the absorption line as a function of Galactic longitude, 
for the sources with high opacities in Fig.~\ref{tass} (filled stars). 
We added the last absorption lines from the spectra of the three sources 
which have no additional line at higher negative velocities and the 
source G90.23, one of the high velocity sources in the right part of 
Fig.~\ref{tauvexp} (open stars). We believe this source belongs to 
this group as well,  because for a high velocity wind the $\tau$ is too
big, and for the dissociated material, shocked or unshocked, the
expansion velocity would be too high.

Apparently there is a strong relation between the radial velocities and the 
Galactic longitude. There are only two sources which do not follow
this relation. One of those exceptions is the source
G92.68 located close to the upper left corner in Fig.~\ref{tass}, which
is of course local based on the low negative radial velocity and should 
not show the spiral shock in the Perseus arm at all. It has
one of the lowest
expansion velocities (0.8~km/s) which is comparable to the uncertainty in our data. So this
strong absorption line is most likely due to a dense \ion{H}{i} cloud
in the foreground. The second exception is the source G132.16 in the 
lower right area of Fig.~\ref{tass}. The $\tau$ is too high to be explained by
the expanding shocked neutral hydrogen layer, but as mentioned in Sect.
4.1, there are rare occasions where the shell of dissociated material
outside the shock front has a systematic velocity of 2-6~km/s for blister
\ion{H}{ii} regions. These shells have higher \ion{H}{i} column densities
and lower spin temperatures so we would expect higher opacities. 
Additionally G132.16 shows another strong absorption line in its spectrum 
at about -40~km/s which would fit nicely to the spiral shock model.

There are 4 sources in our sample which show two additional absorption lines
in their spectra instead of one. G97.51 shows one line with the characteristics of the
spiral shock and the expansion velocity of the shocked neutral hydrogen is
apparently high enough so that both absorption lines are well separated.
The sources G85.25, G93.86, and G105.62 all have one high and one low velocity
component both of which have rather low opacities. We believe that the
higher velocity component is the result of absorption by the shocked
neutral hydrogen while the component with the lower velocity was produced
by the unshocked hydrogen. This would indicate that all three are blister
\ion{H}{ii} regions.

Finally, two of the observed \ion{H}{ii} regions, G75.77+0.35 and
G75.83+0.40, have a very high $v_{exp}$ and low $\tau$. Since both of
these sources are very close together on the sky (only 5$\arcmin$ apart), 
one possibility is that we are seeing unassociated blueshifted foreground 
Galactic gas. We think this is unlikely though since the observed 
absorption line is seen at different velocities in each case 
($v_{rad} = -27.7$~km/s for G75.77 and $v_{rad} = -29.7$~km/s for G75.83). 
Instead we believe we are observing
a fast neutral outflow from these objects, perhaps formed from a
recombining ionized wind. These two objects are by far the most
energetic \ion{H}{ii} regions in our sample, and previous $^{12}$CO
observations have detected high velocity molecular gas, with $v \sim
10 - 50$ km/s, towards these objects \citep{she96}. Models of magnetic
fields in \ion{H}{ii} regions suggest that for certain magnetic field
alignments fast neutral winds ($v \geq 20$ km/s) can form from recombining
ionized flows that have broken out of the molecular cloud
\citep{fra89}. In addition, a fast neutral outflow with $v \leq
100$~km/s has been observed in the high-mass star forming region DR 21
\citep{rus92}. From this, we conclude that we are seeing in absorption
the blueshifted part of a fast neutral outflow coming from these
objects although the linewidth is surprisingly narrow in both
cases. The velocity shift in this case is the outflow velocity
projected along the line of sight.

\subsection{Where is the receding shell?}
 
We identified most of the additional absorption lines as the absorption
of the \ion{H}{ii} region's radio continuum emission by the expanding layer
of shocked neutral hydrogen around the ionized gas. Now we have to discuss
the receding parts of these symmetric layers. By utilizing eqns. (1) and
(2) we can calculate the amplitude ratio of the absorption line of the
approaching shell to the emission line of the receding shell, which
results in 
\begin{displaymath}
\left|\frac{T_B(v_d-v_{exp})}{T_B(v_d+v_{exp})}\right|=\frac{|T_S - T_C|}{T_S}. 
\end{displaymath}
This ratio would be 1 if the radio continuum temperature is twice
the spin temperature. For the compact \ion{H}{ii} regions we should expect
radio continuum brightness temperatures of up to 1000~K, maybe even 2000~K, 
at 1420~MHz, while the spin temperature should be around a few 100~K up
to a maximum of 1000~K. Since most sources show rather complex
foreground absorption spectra it is very difficult to decide if
a structure is an emission line superposed on the foreground
absorption or just the lack of absorption at this position. 
In four cases, G76.15, G76.19, G93.86, and G114.07, we may see
the receding shell in emission. All of these
sources show a possible line at about the expected radial velocity 
$v_d+v_{exp}$. We get
amplitude ratios between the absorption line of the approaching shell
and the emission line of the receding shell of about 3:1 for
G76.15 and G114.07 and about 3:2 for G76.19 and G93.86. This would translate
to $T_c$ to $T_s$ ratios of 4 and 2.5 respectively. 

For those objects where we do not see the emission line one can calculate
the $T_c$ to $T_s$ ratio required to make the line undetectable.
The highest 
$T_c$ to $T_s$ ratios we derive are for G80.35 and G85.25. These sources
show neither emission nor absorption at the expected radial velocity
for the receding shell and we need ratios of around 6 so that the amplitude
of the receding shell is hidden in the noise. This is not a problem for
the typical values for $T_c$ and $T_s$ we discussed above.

\subsection{G76.15-0.29, a special case?}
 
For none of the sources in our sample we would expect to observe either 
of the neutral layers around them in emission because these sources are
unresolved and thus this emission should be buried within the absorption
lines. Especially the thin shocked layer should be impossible
to detect in our data. However, around the source G76.15, which is slightly extended
in our data, we found an incomplete shell of \ion{H}{i} emission centered 
on its recombination line velocity (see Fig.\ref{hishell1}). This \ion{H}{i}
shell has a u-shaped structure open to the north-east. In the north-west
and the south-east it seems to be located right outside the \ion{H}{ii}
region while there is a gap between them in the south-west. We derived
emission profiles for this structures at different radii from
the center of the \ion{H}{ii} region going to the north-west
(see Fig.\ref{hishell2}). The separation of the two emission peaks
indicate an expansion velocity of 5 - 6 km/s. The existence of the emission gap 
between the \ion{H}{ii} region and the \ion{H}{i} shell in the south-west
demonstrates that the \ion{H}{i} shell cannot be the layer of shocked
neutral hydrogen, since this layer should be located right outside the
ionization front. Hence, this expanding shell must be the layer
of unshocked dissociated molecular material. The only explanation
for the systematic velocity difference of 5 - 6~km/s is that the \ion{H}{ii}
region is not completely surrounded by molecular material. Thus
we can expect flows of neutral hydrogen towards low density areas
giving the material a velocity along the line of sight. G76.15 must
also be a blister \ion{H}{ii} region.

\section{Summary}

We developed a background filtering technique that allows the detection of 
small-scale positive and negative features within a data set. The filtering 
was applied to CGPS \ion{H}{i} observations towards 26 compact \ion{H}{ii} 
regions in order to obtain absorption and emission spectra.

In all but three of the spectra we observe absorption lines at more negative 
velocities than would be expected from the absorption of continuum emission 
by foreground material.  A simple model of an expanding \ion{H}{ii} region 
with surrounding shocked and unshocked \ion{H}{i} zones was developed to 
interpret the observations. For the majority of the observations, the physical 
properties (expansion velocity, column density) of the shells derived from 
the absorption spectra are consistent with them being caused by shocked 
expanding shells of \ion{H}{i}.

Some of the additional lines we observe are caused by a large scale Galactic 
shock associated with the Perseus arm of the Galaxy. Evidence for this comes 
from the way in which the velocity of the observed absorption features vary 
with longitude.

Finally, two of the observed \ion{H}{ii} regions show evidence for high 
velocity neutral flows such as the flows observed in emission towards DR~21.

This study has shown how absorption spectra can be used to examine neutral 
material surrounding \ion{H}{ii} regions even if the \ion{H}{ii} regions are 
unresolved. The techniques that have been developed should prove to be useful 
as radio observations of the Galactic plane at $1'$ resolution are extended 
beyond the CGPS region to cover the majority of the plane.

\begin{acknowledgements}
The Dominion Radio Astrophysical Observatory is a National Facility
operated by the National Research Council.  The Canadian Galactic
Plane Survey is a Canadian project with international partners, and is
supported by the Natural Sciences and Engineering Research Council
(NSERC).
\end{acknowledgements}


\begin{figure*}[h]
   \caption{\ion{H}{i} absorption spectra of six compact \ion{H}{ii}
      regions. These spectra were taken at the peak position of the
      continuum source. Radial velocities are indicated by a dashed line.
      Upper lines show the emission profile at the position of
      the \ion{H}{ii} regions taken from $T_{large}$ and the lower line 
      the absorption profile taken from $T_{small}$. Arrows indicate
      the absorption features of interest.}
   \label{firstspec}
\end{figure*}
\begin{figure*}
   \caption{see Fig. \ref{firstspec}}
\end{figure*}
\begin{figure*}
   \caption{see Fig. \ref{firstspec}, positive structures in the
   absorption profiles of G96.29 and G97.51 are filaments coincidentally
   at this position and not related to the \ion{H}{ii} regions}
\end{figure*}
\begin{figure*}
   \caption{see Fig. \ref{firstspec}, positive structures in the
   absorption profiles of G112.23 and G115.79 are filaments coincidentally
   at this position and not related to the \ion{H}{ii} regions}
\end{figure*}
\begin{figure*}
   \caption{see Fig. \ref{firstspec}}
   \label{lastspec}
\end{figure*}

\begin{figure*}
   \caption{Left: A sketch of a compact \ion{H}{ii} region (H$^+$) with an 
    expanding layer of shocked neutral hydrogen (H$^*$) as 
    described in Sect. 4.1. The dissociation front (DF) and ionization
    front (IF) are indicated, as is the layer of the unshocked \ion{H}{i}
    (H). Right: A diagram of the radial velocity
    as a function of the distance from the Sun at a longitude of $76\degr$ 
    for a flat rotation model with $v_\odot$ = 220~km/s and 
    $R_\odot$ = 8.5~kpc.
    Velocity shifts of an expanding shell with $v_{exp}
    = 10$~km/s for an \ion{H}{ii} region with a radial velocity of 
    $v_d=-40$~km/s are indicated.}
   \label{model}
\end{figure*}

\begin{figure*}
   \caption{Four example absorption profiles with the fitted
   Gaussians. The histograms represent
   the observed brightness temperature distribution and the solid
   line the fitted Gaussians.}
   \label{gauss}
\end{figure*}

\begin{figure*}
   \caption{Diagram with calulated minimum $\tau$ values for all
   \ion{H}{ii} regions as a function of the expansion velocity.
   Three groups of sources with possible different origin for the
   additional absorption line are marked.}
   \label{tauvexp}
\end{figure*}

\begin{figure*}
   \caption{Diagram of the radial velocity as a function of Galactic longitude
      for sources with high calculated $\tau$ values. Filled stars
      represent sources which are located in the box marked ``spiral shocks''
      in Fig.~\ref{tauvexp}. Open stars represent the last absorption line of
      those sources which have no additional absorption line at a velocity
      $v < v_{LSR}$ and the source G90.24.}
   \label{tass}
\end{figure*}

\begin{figure*}
   \caption{The \ion{H}{i} shell found in emission around the \ion{H}{ii}
   region G76.15 integrated between -22 and -30~km/s.}
   \label{hishell1}
\end{figure*}

\begin{figure*}
   \caption{Emission profiles of the \ion{H}{i} shell around G76.15 taken
   at different radii.}
   \label{hishell2}
\end{figure*}


\begin{thebibliography}{}

\bibitem[Brand \& Blitz(1993)]{brand} Brand J., Blitz L., 1993, A\&A 275, 67

\bibitem[Brunt \& Kerton(2002)]{brunt} Brunt C.M., Kerton C., 2002, \apj, 
   submitted

\bibitem[Bronfman et al.(1996)]{bronfman} Bronfman L., Nyman L.-A., May J.,
   1996, A\&AS 115, 81

\bibitem[Dewdney \& Roger(1982)]{dr82} Dewdney P. E. \& Roger R. S., 
   1982, \apj, 235, 564

\bibitem[Franco, Tenorio-Tagle, \& Bodenheimer(1989)]{fra89} Franco, J., 
   Tenorio-Tagle, G., \& Bodenheimer, P. 1989, RMxAA, 18, 65
   
\bibitem[Garay et al.(1993)]{garay} Garay G., Rodriguez L.F., Moran J.M.,
   Churchwell E., 1993, ApJ 418, 368

\bibitem[G\'omez et al.(1998)]{gom98} G\'omez Y., Lebron M., Rodriguez L.F.,
     Garay G., Lizano S., Escalante V., Canto J., 1998, \apj, 503, 297

\bibitem[Higgs(2000)]{hig00} Higgs, L. A. 2000, in Imaging at Radio
through Submillimeter Wavelengths, eds. J.G. Mangum \& S.J.E. Radford,
ASP Conf. Ser. 217, 259

\bibitem[Hill \& Hollenbach(1978)]{hh78}Hill J. K. \& Hollenbach D. J., 
   1978, \apj, 225, 390

\bibitem[Kuchar \&  Bania(1990)]{kb90}Kuchar, T. A. \& Bania, T. M. 1990, \apj, 352, 192

\bibitem[Kuchar \&  Bania(1994)]{kb94}Kuchar, T. A. \& Bania,
T. M. 1994, \apj, 436, 117

\bibitem[Kurtz et al.(1994)]{kurtz} Kurtz S., Churchwell E., Wood D.O.S.,
   1994, ApJS 91, 659

\bibitem[Landecker et al.(2000)]{lan00}Landecker, T. L. et al. 2000,
\aaps, 145, 509

\bibitem[Lebr\'on et al.(2001)]{le01} Lebr\'on, M. E., Rodriguez, L. F., \& 
   Lizano, S. 2001, \apj, 560 , 806

\bibitem[Lockman(1989)]{lockman} Lockman F.J., 1989, ApJS 71, 469

\bibitem[McClure-Griffiths et al.(2001)]{mcc01}McClure-Griffiths,
N. M. et al. 2001, \apj, 551, 394 

\bibitem[Normandeau(1999)]{nor99} Normandeau, M. 1999, \aj, 117, 2440

\bibitem[Richards et al.(1987)]{richards} Richards P.J., Little L.T., 
   Toriseva M., Heaton B.D., 1987, MNRAS 228, 43
      
\bibitem[Roberts(1972)]{roberts} Roberts W.W., 1972, \apj, 173, 259

\bibitem[Roberts, Dickel, \& Goss(1997)]{rob97}Roberts, D. A.,
Dickel, H. R., \& Goss, W. M. 1997, \apj, 476, 209

\bibitem[Roger(1982)]{rog82} Roger, R. S. 1982, in Regions of Recent Star Formation, eds. R. S. Roger \& P. 
   E. Dewdney, (Dordrecht; Reidel), 167

\bibitem[Roger \& Dewdney(1992)]{rog92} Roger, R. S. \& Dewdney,
P. E. 1992, \apj, 385, 536

\bibitem[Roger \& Irwin(1982)]{ri82} Roger, R. S. \& Irwin,
J. A. 1982, \apj, 256, 127

\bibitem[Roger \& Pedlar(1981)]{rog81} Roger, R. S. \& Pedlar,
A. 1981, \aa, 94, 238

\bibitem[Russell et al.(1992)]{rus92} Russell, A. P. G., Bally, J., Padman, R., 
   \& Hills, R. E. 1992, \apj, 387, 219

\bibitem[Shepherd \& Churchwell(1996)]{she96} Shepherd, D. S. \& Churchwell, E. 
   \apj, 457, 267

\bibitem[Sofue \& Reich(1979)]{sofue} Sofue, Y., Reich, W., 1979, 
          \aaps, 38, 251

\bibitem[Taylor(1999)]{tay99}Taylor, A. R. 1999, in New Perspectives
   on the Interstellar Medium, eds. A. R. Taylor, T. L. Landecker, and
   G. Joncas, ASP Conf. Ser. 168, 3

\bibitem[Timmermann et al.(1996)]{tim96}Timmermann R., Bertoldi F., 
   Wright C. M., Drapatz S., Draine B.T., Haser L., \& Sternberg A. 
   1996, \aap, 315, L281

\bibitem[van den Ancker, Tielens, \& Wesselius(2000)]{vda00} 
   van den Ancker M. E., Tielens A. G. G. M., Wesselius P. R. 2000,
   \aap, 358, 1035

\bibitem[van der Werf \& Goss(1989)]{vdw89}van def Werf, P. P. \&
   Goss, W. M. 1989, \aap, 224, 209

\bibitem[van der Werf \& Goss(1990)]{vdw90}van def Werf, P. P. \&
   Goss, W. M. 1990, \aap, 238, 296

\bibitem[Wendker \& Wrigge(1996)]{wen96} Wendker, H. J. \& Wrigge,
   M. 1996, \aap, 305, 592

\bibitem[Wood \& Churchwell(1988)]{wood} Wood D.O.S. \& Churchwell E., 1988,
   ApJS 69, 831

\bibitem[Wouterloot \& Brand(1989)]{wouterloot} Wouterloot J.G.A., Brand J., 
   1993, A\&AS 80, 149
    
\end{thebibliography}
\end{document}